\documentclass[preprint,aps,pre,superscriptaddress,showpacs,floatfix]{revtex4}

\usepackage[dvips]{graphicx}
\usepackage{amsmath}
\usepackage{epsfig}

\begin{document}
\preprint{draft}
\title{Self-averaging in random systems - liability or asset?}
\author{Avishay Efrat}
\affiliation{Holon Academic Institute of Technology, Holon 58102, Israel}
\author{Moshe Schwartz}
\affiliation{School of Physics and Astronomy, Tel-Aviv University, Ramat-Aviv, Tel-Aviv 69978, Israel}
\date{\today}
\begin{abstract}
The study of quenched random systems is facilitated by the idea that the ensemble averages describe the thermal averages for any specific realization of the couplings, provided the system is large enough. Careful examination suggests that this idea might have a follow, when the correlation length becomes of the order of the size of the system. We find certain bound quantities are not self-averaging when the correlation length becomes of the order of the size of the system. This suggests that the strength of self-averaging, expressed in terms of properly chosen signal to noise ratios, may serve to identify phase boundaries. This is demonstrated by using such signal to noise ratios to identify the boundary of the ferromagnetic phase and compare the findings with more traditional measures.
\end{abstract}
\maketitle

Quenched random systems like random field systems, random bond systems, spin glasses etc. are known to be extremely difficult, because of the necessity to perform quenched averages. In fact, it has to be realized, of course, that this necessity, is not of physical origin but rather of mathematical convenience, since a given chunk of matter that is measured experimentally has a given single realization of the disorder. The idea of self-averaging allowing ensemble averaging is based on the fact that we are dealing with very large systems \cite{b59}. It is argued that the system can be broken up into subsystems large enough to be considered independent of each other. This basic assumption indicates that phenomena connected with breakdown of self-averaging will appear, as the correlation length will become of the order of the linear size of the system. Indeed, such behavior was first observed by Dayan et al \cite{dsy93}, using a technique first suggested by Berker and Ostlund \cite{bo79}. The subject of breakdown of self-averaging was treated later in a number of papers devoted just to that phenomenon \cite{wd95,ah96,wd98,mf06}. Obviously, lack of self-averaging results in severe difficulties but the fact that it is connected with the divergence of the correlation length suggests that it can provide, perhaps, an independent measure to distinguish between the disordered and the ordered phase. In this article we show that this is indeed the case.

To be specific consider the random field Ising model described by the Hamiltonian
\begin{equation}
{\cal H}=-J\sum_{<i,j>}\sigma_i\sigma_j-\sum_i h_i\sigma_i.
\label{EqFieldHamJ}
\end{equation}
The pair $<i,j>$ denotes a nearest-neighbor pair on a cubic lattice. The field configuration is assumed to be governed by the distribution,
\begin{equation}
P\{h\} = \prod_i \frac{1}{\sqrt{2\pi h^2}}\textrm{exp}\left(-\frac{{h_i}^2}{2h^2}\right).
\label{EqGaussDist}
\end{equation}
We take $J$ and $h$ to have the dimensions of energy.

Consider now some quantity $\alpha$ and its ensemble average, $\alpha_E$. Define next the variance of $\alpha$ within the ensemble,
\begin{equation}
\sigma_\alpha \equiv \{[\alpha^2]-{\alpha_E}^2 \}^\frac{1}{2},
\label{EqVarAlpha}
\end{equation}
where $[\ldots]$ denotes ensemble average. The parameter $\gamma$ describing the strength of self-averaging in the system (or rather its weakness) is defined as the signal to noise ratio
\begin{equation}
\gamma_\alpha \equiv \frac{|\alpha_E|}{\sigma_\alpha}.
\label{EqDefgamma}
\end{equation}
It is clear that for the disordered phase $\gamma$ is infinite in the infinite system. Our former discussion suggests that possible breakdown of self-averaging may be traced back to situations where the correlation length, $\xi$, becomes of the order of the linear size of the system. The usual state of affairs is that the correlation length diverges just on the boundary between the ordered and disordered phase and consequently any break down of self-averaging may be observed only in the vicinity of the boundary \cite{wd95,ah96,wd98}. In the random field case the situation is quite different. The correlation length is of the linear size of the system everywhere in the ordered phase and not just at the boundary of the phase \cite{dsy93}.

\begin{figure}[!]
\includegraphics[width=.48\textwidth]{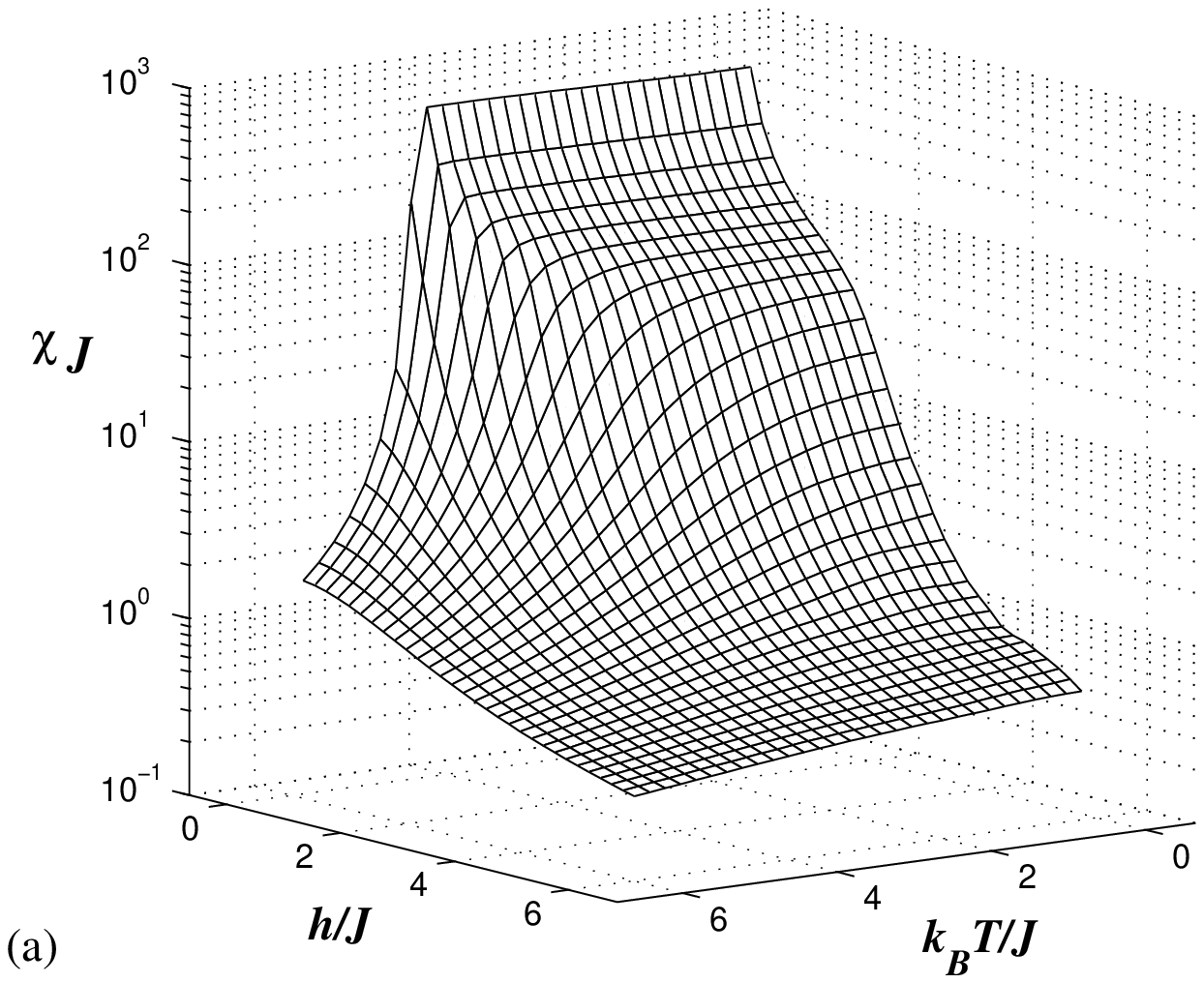}
\includegraphics[width=.48\textwidth]{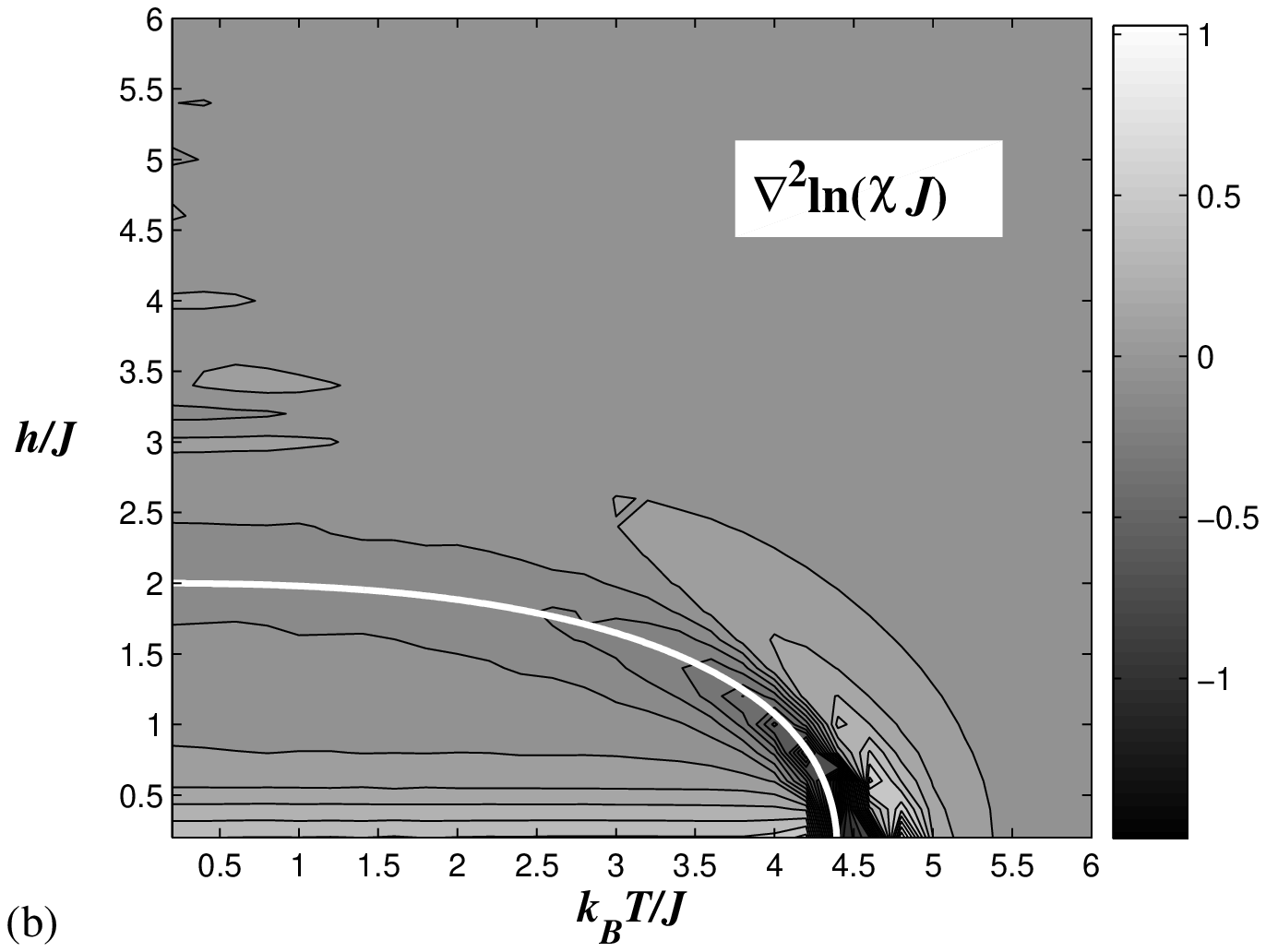}
\caption{\label{Fig1} 
\footnotesize{The logarithm of the susceptibility (a), and its Laplacian (b) everywhere, in dimensionless units, as a function of the dimensionless temperature, $k_BT/J$, and the dimensionless strength of the random field, $h/J$. In (b), a pseudo-critical line of the finite $32\times32\times32$ system is represented by the white line marking the dark ridge separating the two brighter plateaus.}}
\end{figure}

To understand the possible different behaviors of $\gamma$ consider first the case where $\alpha$ is the free energy. Suppose we break the system into mutually non-interacting sub systems of linear size $L' \ll L$ and evaluate the free energy of the system as the sum of the free energies of the sub systems. This generates a relative error in the free energy that is of the order of $1/L'$ regardless of the size of the correlation length. This suggests that nothing spectacular will happen to the signal to noise ratio of the free energy when we cross from the ordered to the disordered phase. To see if this is true we need to know, first, where is the ordered phase boundary. In Fig. \ref{Fig1}a we give the logarithm of the susceptibility. The ordered phase is presented as an upper plateau. In Fig. \ref{Fig1}b we give the Laplacian of the logarithm of the susceptibility with respect to the coordinates given in Fig. \ref{Fig1}a. The white line drown in Fig. \ref{Fig1}b is where the absolute value of the Laplacian, which is related to the curvature, is maximal and is viewed as the phase boundary. Our result for the boundary is consistent with phase diagrams (or usually parts of it) obtained by others \cite{rgsl85,o86,nb96,fh9698,sbmcb97,as97,hn99,mnc00,hy01,mf02}.

\begin{figure}[!]
\includegraphics[width=.48\textwidth]{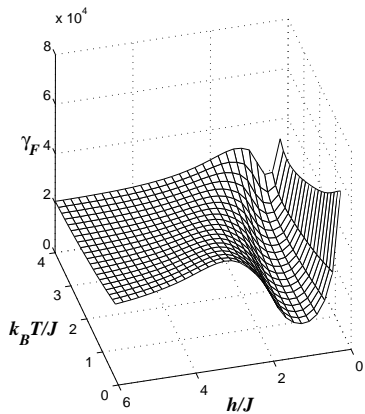}
\caption{\label{Fig2} 
\footnotesize{The signal to noise ratio of the free energy of a system of size $32\times32\times32$. Everywhere it is of the order of $10^4$.}}
\end{figure}

\begin{figure}[!]
\includegraphics[width=.48\textwidth]{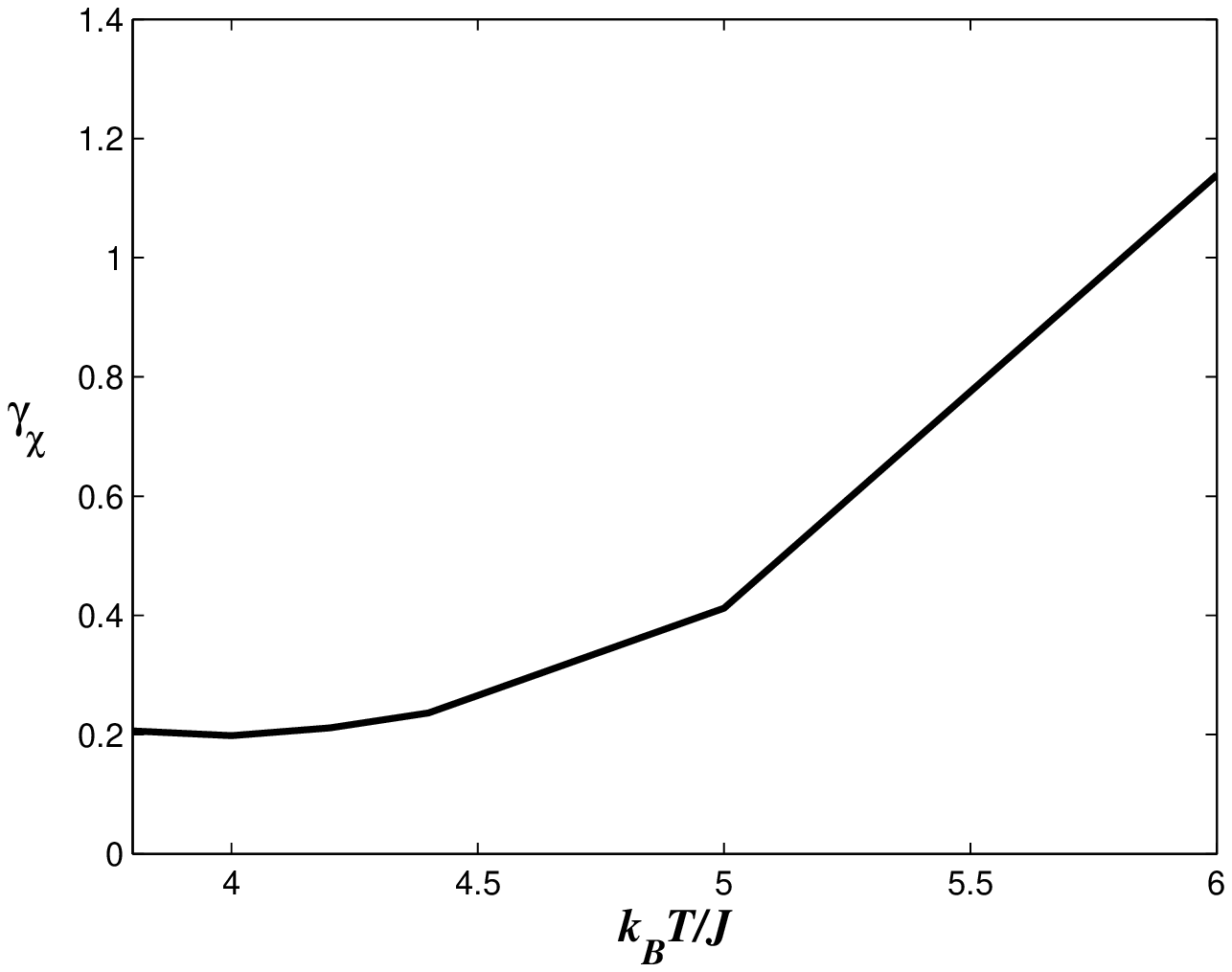}
\caption{\label{Fig3} 
\footnotesize{The signal to noise ratio in the susceptibility for a system of $32\times32\times32$ and for a given ratio of $k_BT/h=2$, obtained from data presented in Ref. \cite{dsy93}. The critical dimensionless temperature is $k_BT/J=4.2$.}}
\end{figure}

\begin{figure}[!]
\includegraphics[width=.48\textwidth]{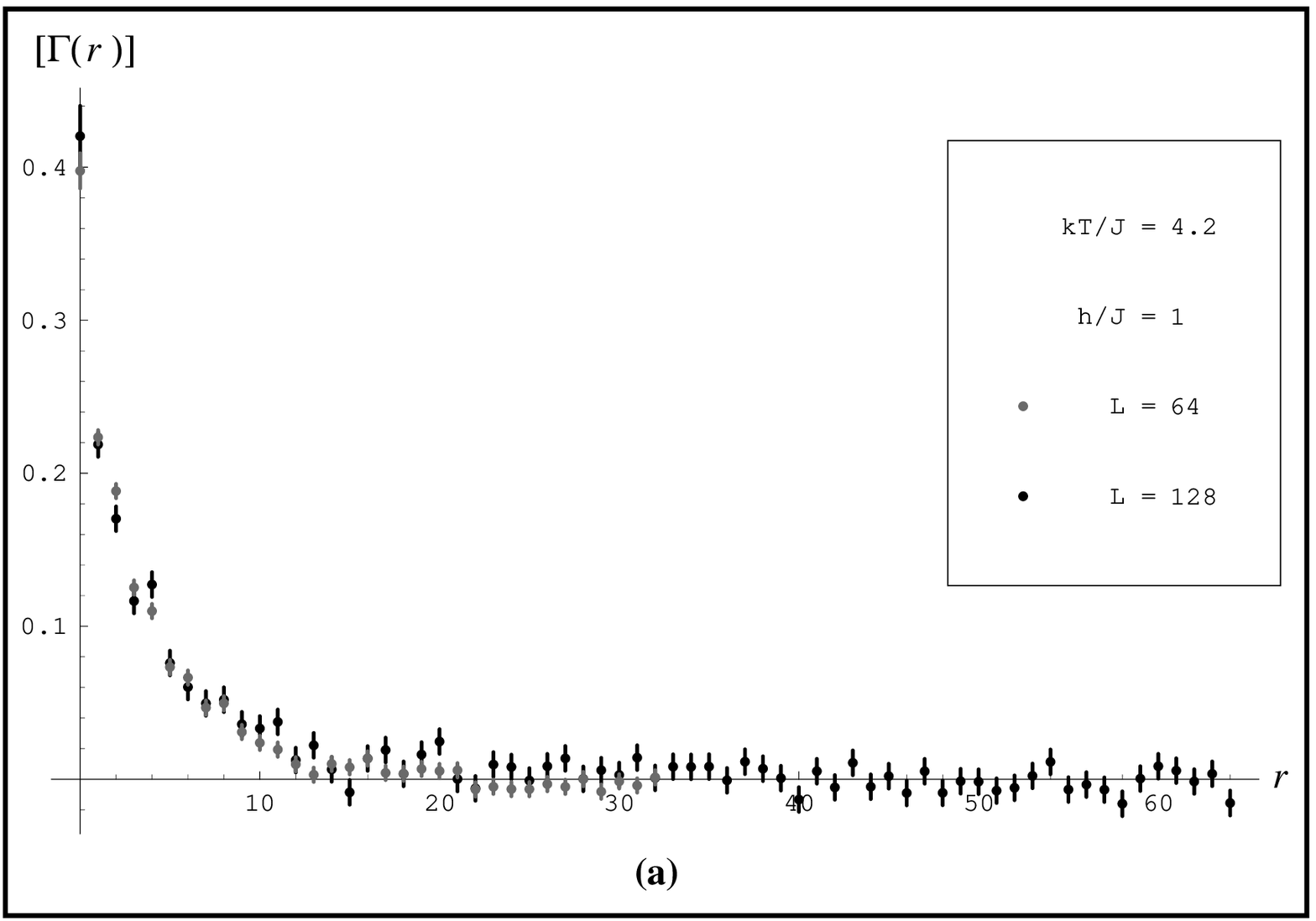}
\includegraphics[width=.48\textwidth]{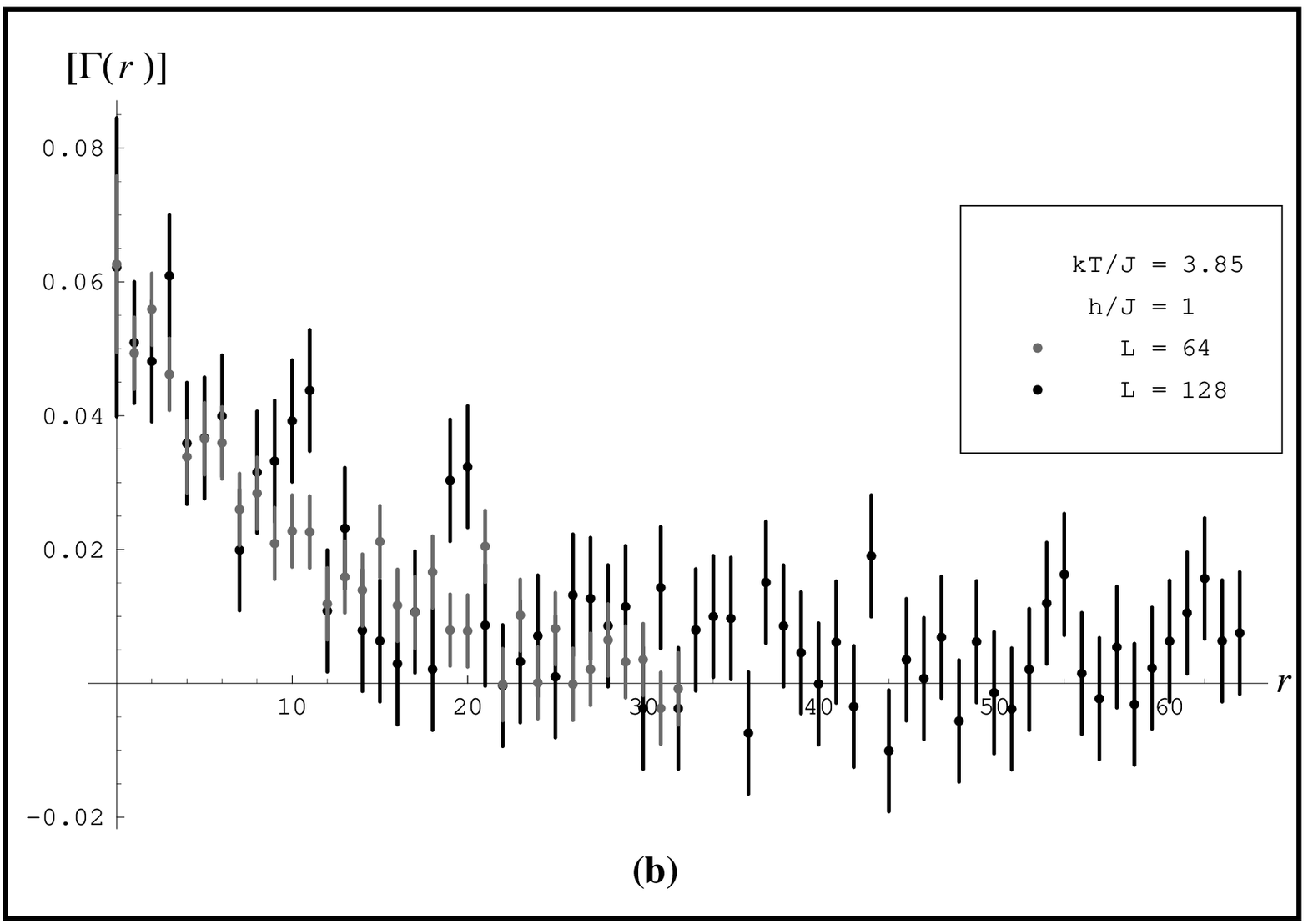}
\includegraphics[width=.48\textwidth]{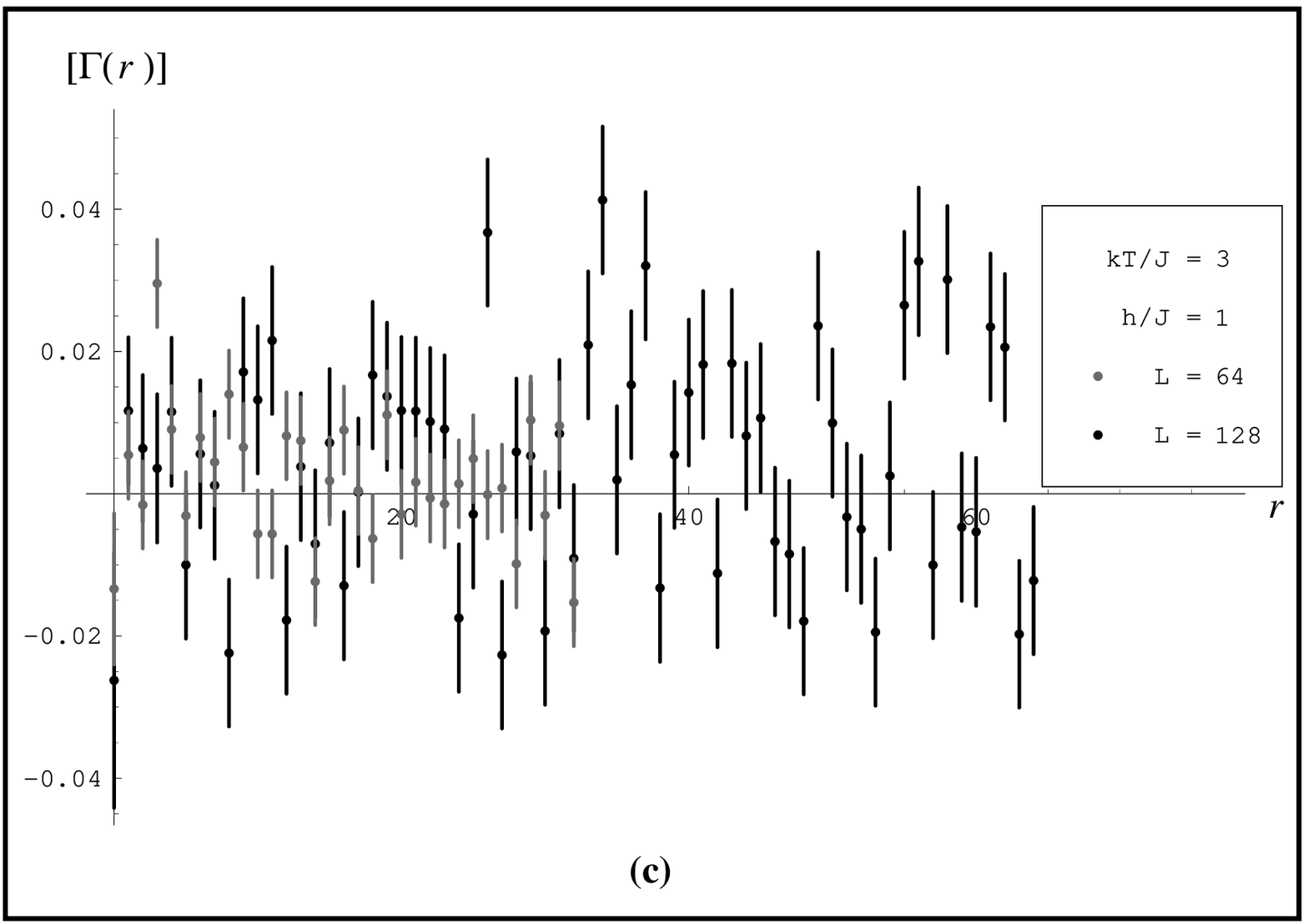}
\caption{\label{Fig4} 
\footnotesize{(Taken from Ref. \cite{es03}) The average spin-spin correlation function, $\Gamma_E$, is shown as a function of the distance, $r$, taken along the main axes of the lattice and measured in units of lattice constant. The external field, $H$, is zero while the standard deviation of the RF is $h/J=1$. As indicated on each figure, the two different levels of gray scaling correspond to systems of different linear size, $L$, indicating the size independence of
$\Gamma_E$ for these temperatures. The points of the larger system, with $L=128$, appear to be more scattered since it is averaged only over $3500$ realizations, while the smaller, $L=64$, system is averaged over $10000$ realizations. The different figures correspond to different temperatures. Note the broadening of $\Gamma_E$ as the temperature is reduced [(a) and (b)] towards entering the ordered phase, at about $k_BT/J=3.8$, untill, due to noise [as in (c)], no trend can be detected.}}
\end{figure}

In Fig. \ref{Fig2} we give the signal to noise ratio of the free energy. Nothing spectacular happens in the vicinity of the phase boundary. The signal to noise ratio in the susceptibility is another matter. It remains small anywhere in the ordered phase. For the sake of completeness we give in Fig. \ref{Fig3} the temperature dependence of the signal to noise ratio associated with the susceptibility as obtained from the results of Ref. \cite{dsy93}. 
If only signal to noise ratios in divergent quantities showed some interesting behavior it would not have been very interesting as indeed it would have provided no additional information to that obtained by calculating the average itself. It is thus of interest to consider physical quantities that do not diverge in the ordered phase or on its boundary. Our choice is motivated by our previous work on real space renormalization of the three dimensional random field Ising model (RFIM) \cite{es03}. In that article we have obtained, for the first time, the "connected" spin-spin correlation,
\begin{equation}
\Gamma_E(r) \equiv [\Gamma(r)] \equiv [\langle\sigma(0)\sigma(r)\rangle-\langle\sigma(0)\rangle\langle\sigma(r)\rangle].
\label{EqDefGamma}
\end{equation}
In Fig. \ref{Fig4} we present $\Gamma_E$ for the case where the strength of the dimensionless random field is $h/J=1$, for linear sizes of $L=64$ and $L=128$ and three different temperatures. It is clear that as temperature is lowered the noisiness increases up to the lowest temperature (Fig. \ref{Fig4}c), where even the trend of decay as a function of $r$ cannot be observed any more. Since $\Gamma_E$ must be a decaying function of the distance this implies that self-averaging must be broken in Fig. \ref{Fig4}c.

An attempt to measure $\Gamma(r)$ can be made, perhaps, along the following lines. An experiment is carried out on a given chunk of matter with some fixed position dependent interactions, so that the measured quantities are spatial averages on that system rather than ensemble averages. The spin-spin $q$ dependent correlation function, $S(\vec{q})$, can be obtained experimentally from neutron scattering \cite{b85,bkj85,sb96,sbf9900,ymkybsfa05}, making the proper allowance that the measurement is done on a dilute anti-ferromagnet in an external constant field rather than on a ferromagnet in the presence of a random magnetic field. The question now is whether the "connected" correlation, $\chi(\vec{q})$, can be unpicked from $S(\vec{q})$. To do that a relation between the two has to be assumed \cite{sbf9900,ymkybsfa05}. The spatial rather than the ensemble average of the $\Gamma_S(\vec{r})$ is given by the Fourier transform of $\chi(\vec{q})$,
\begin{equation}
\Gamma_S(\vec{r}) = \int d\vec{q} \chi(\vec{q}) \exp(i\vec{q}\cdot\vec{r}).
\label{EqGammaS}
\end{equation}
The evaluation of $\Gamma_S(\vec{r})$ has to be repeated for a large enough number of systems in order to obtain reasonable ensemble averages. We will concentrate in the following on the simpler quantity $\Gamma(0)$ and make our point by obtaining the corresponding $\gamma$ from real space renormalization of the random field Hamiltonian (\ref{EqFieldHamJ}). First, however, let us obtain a crude picture of what to expect. Consider some local bounded quantity, $\Gamma_i$, such that the ensemble average of $\Gamma_i$ is $\Gamma_E(0)$. Consider next the spatial average
\begin{equation}
\Gamma_S(0) = \frac{1}{N}\sum_i \Gamma_i,
\label{EqGammaS0}
\end{equation}
where $N$ is the total number of sites in the system. As long as self-averaging is not broken $\Gamma_S(0)$ is expected to "equal" its ensemble average, $\Gamma_E(0)$. (By "equal" we actually mean that the corresponding signal to noise will be very large.) We can get an idea of the variance in $\Gamma_S(0)$ by noting that the sum in the numerator on the right hand side of Eq. (\ref{EqGammaS0}) is a sum of correlated random numbers that are correlated over a distance $\xi$. This implies that the corresponding signal to noise, $\gamma_\Gamma$, is given by
\begin{equation}
\gamma_\Gamma \sim \left(\frac{N}{\xi^d}\right)^{\frac{1}{2}}
\label{EqgammaGamma}
\end{equation}
for a $d$ dimensional system. (Indeed, when approaching the boundary between phases from the disordered phase a more accurate exponent should be used reflecting the nature of algebraic decay of correlations within the correlation length. This will replace $d/2$ in Eq. (\ref{EqgammaGamma}) by another (smaller) positive power of $\xi$.) In any case the signal to noise must decrease considerably in the ordered phase compared to its values in the disordered phase. (Note that the free energy, cannot be written in the form given by Eq. (\ref{EqGammaS0}) and therefore the argument given above does not apply to signal to noise ratio of the free energy.) It is clear ,however, from the argument above that relation (\ref{EqgammaGamma}) is of a generic nature and does not depend on the specific quantity used. The ratio of two signal to noise ratios or two variances corresponding to two different quantities will thus not be essentially different in the two phases. 

We describe next our procedure for obtaining the signal to noise ratio of $\Gamma(0)$ numerically, which is based on Casher-Schwartz RSRG \cite{cs78}. As other real space techniques, it provides simple, one step, recursion relations for translational invariant systems that enable the extraction of the critical exponents. For random systems, the recursion relations obtained, using any renormalization scheme, involve the distribution of couplings or, equivalently, all the parameters defining it (e.g. moments, correlations etc.). In this approach, the recursion relations are truncated to obtain relations involving only the mean and the variance and keeping the random couplings independent \cite{sf80,kd81,ab84,aa85}. The method suggested by Berker and Ostlund \cite{bo79} overcomes the most difficult problem arising in the approach described above of ignoring the correlations generated by the renormalization procedure (or projecting them on the variance \cite{sf80}). A given realization is chosen on a finite system. Renormalization is then used to reduce the size of the system to a size where brute force calculation is possible. (The fact that a lot can be learned from considering specific realizations is also stressed in a recent paper on the RFIM by Wu and Machta \cite{wm05}.) After the required thermal averages are obtained for a given configuration of the disorder, the ensemble average is obtained by repeating the procedure for many configurations and averaging. Only thermal averages of functions of the spins surviving the renormalization seem to be obtainable directly. Berker and coworkers \cite{fbm95,fb9697,yb97} were using, however, the chain rule to approximately recover thermodynamic densities of the original system from the renormalized couplings of the reduced system. This enables to obtain averages which involve not only surviving spins, but is limited to thermal averages of spin products which appear in the original Hamiltonian (such as nearest-neighbor pair products). In the following we will see how averages involving non surviving spins such as $\Gamma(r)$ can be calculated \cite{es03}.

We start with a set of $N=L^3$ Ising spins, with $L=2^n$, situated on a three dimensional cubic lattice with periodic boundary conditions. We generate a realization of the distribution (\ref{EqGaussDist}) and then perform the Casher-Schwartz procedure \cite{cs78} $n-1$ times. At each step of the renormalization, the lattice remains cubic while its linear size is reduced by a factor of two. The couplings remain nearest-neighbor but become position dependent and each spin is affected by a renormalized position dependent field. Finally, we are left with a system of $2\times2\times2$ spins on which we perform direct calculations. Consequently we obtain $8$ $\langle\sigma_i\rangle$'s, corresponding to the spins that survive the renormalization procedure. We define a local, position dependent, quantity $\Gamma_i(0)$,
\begin{equation}
\Gamma_i(0) \equiv \frac{1}{\beta h^2}\langle\sigma_i\rangle h_i,
\label{EqGammaI0}
\end{equation}
where $h_i$ is the original field on the site $i$. The actual ensemble averages are performed by repeating the calculations for many realizations of the randomness, summing up the results and dividing by the number of realizations. The statistics is improved by using, instead of the local quantity defined above, its average over the spins surviving the renormalization procedure,
\begin{equation}
\Gamma(0) = \frac{1}{8\beta h^2}\sum_{i=1}^{8}\langle\sigma_i\rangle h_i,
\label{EqGamma0}
\end{equation}
where $i$ runs over the $8$ surviving spins. (Note that the ensemble average of $\Gamma(0)$ is indeed the "connected" self correlation $\Gamma_E(0)$ \cite{es03,ss8586}.) The variance is
\begin{equation}
\sigma_{\Gamma(0)} = \{[\Gamma^2(0)]-{\Gamma_E}^2(0)\}^\frac{1}{2}.
\label{EqSigmaGamma0}
\end{equation}

\begin{figure}[!]
\includegraphics[width=.48\textwidth]{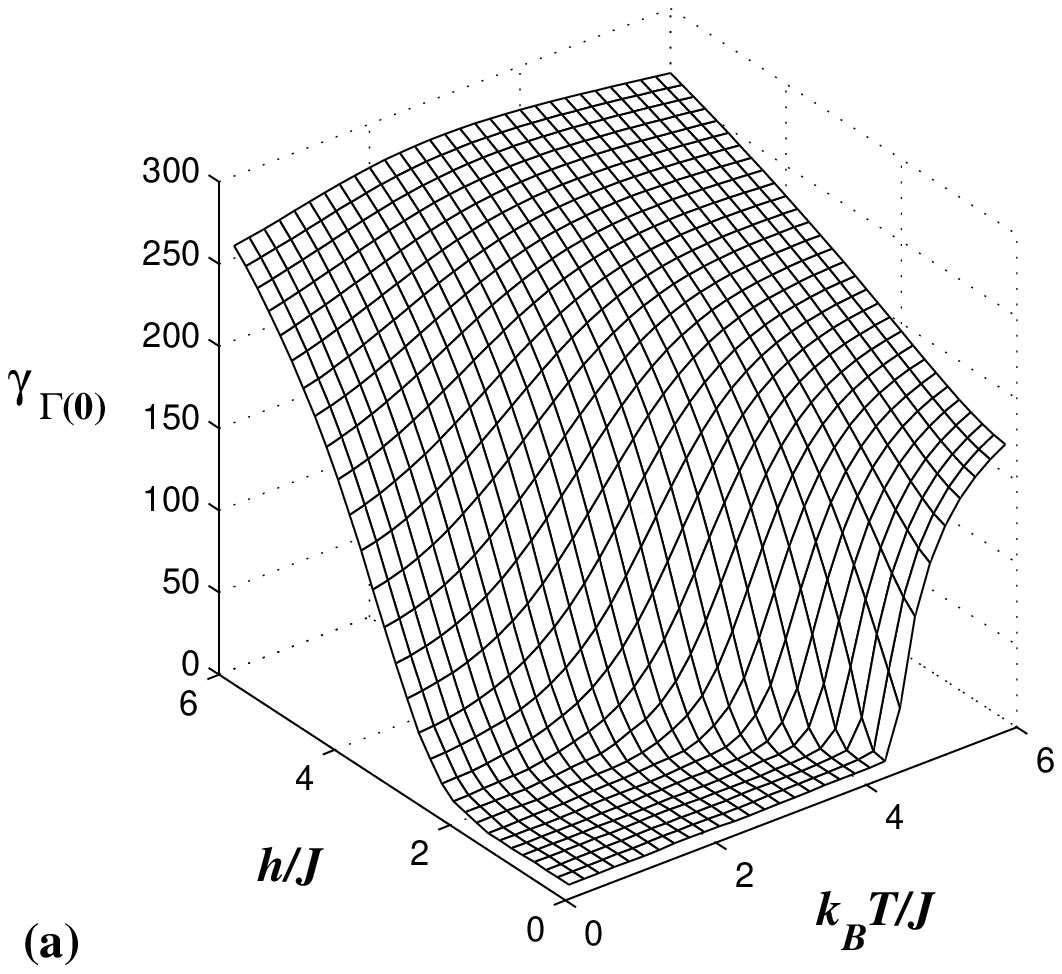}
\includegraphics[width=.48\textwidth]{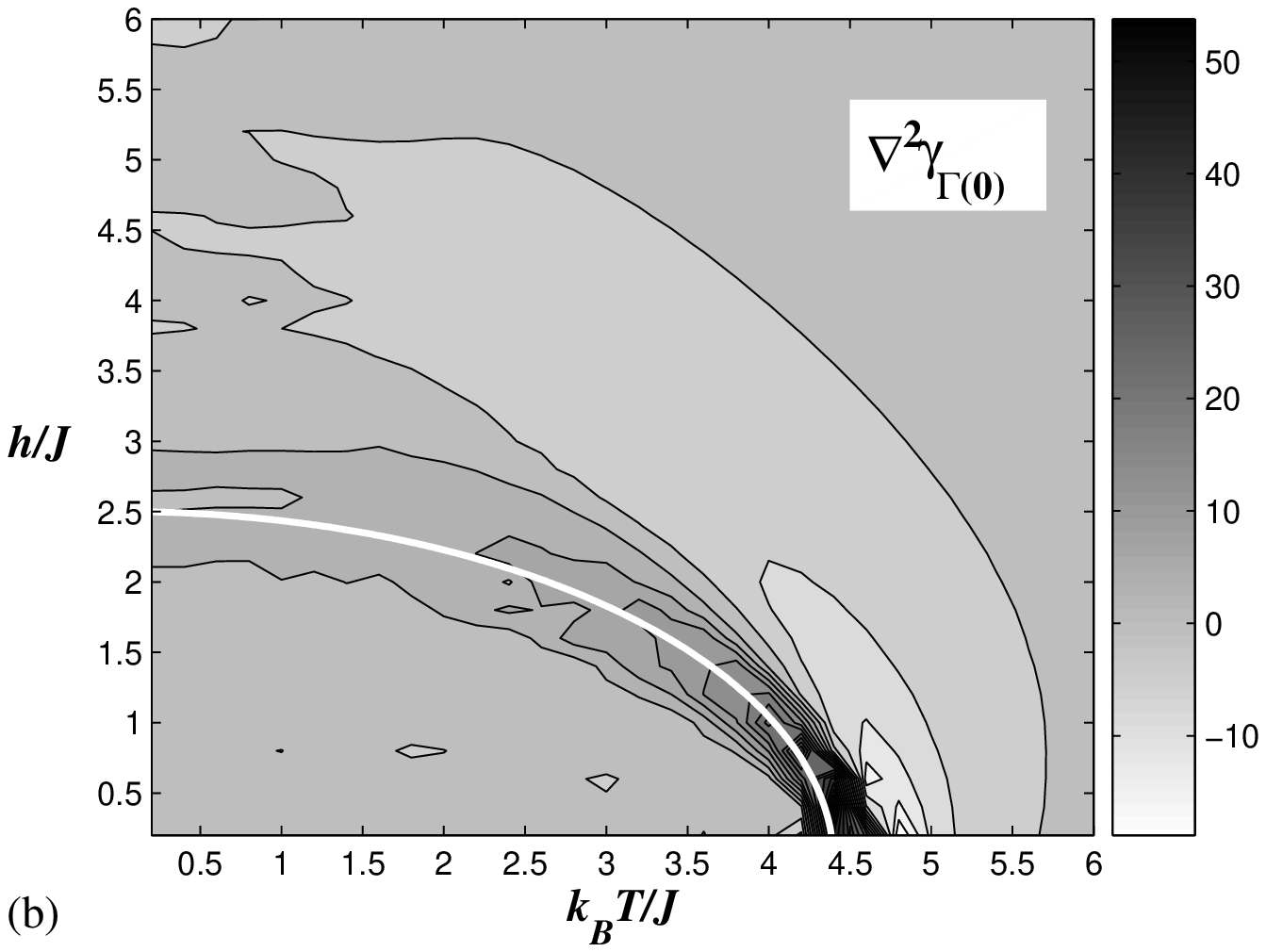}
\caption{\label{Fig5} 
\footnotesize{The signal to noise ratio in $\gamma_{\Gamma(0)}$ (a) and its Laplacian (b) everywhere. The strong reduction in the ordered phase, shown in (a), is evident. As in Fig. \ref{Fig1}, in (b), the white line marks the maximum of the Laplacian and taken as the phase boundary.}}
\end{figure}

\begin{figure}[!]
\includegraphics[width=.48\textwidth]{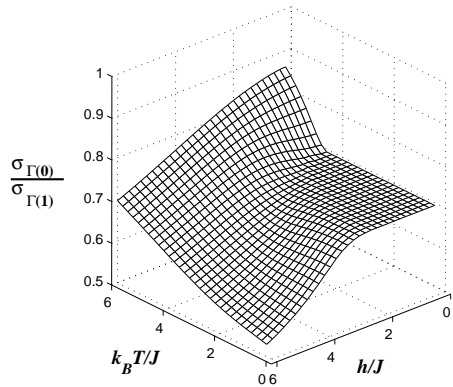}
\caption{\label{Fig6} 
\footnotesize{The ratio of $\sigma_{\Gamma(0)}$ and $\sigma_{\Gamma(1)}$. This does not show any pronounced dependence on the correlation length.}}
\end{figure}

The signal to noise in $\Gamma(0)$ is presented in Fig. \ref{Fig5}. The strong reduction in the ordered phase is very clear.
We proceed now to check relation (\ref{EqgammaGamma}), which is supposed to be generic. We define first the quantity
\begin{equation}
\Gamma(1) \equiv \frac{1}{48\beta h^2}\sum_{i=1}^{8}\sum_{n(i)=1}^{6}\langle\sigma_i\rangle h_{n(i)},
\label{EqGamma1}
\end{equation}
where $n(i)$ runs over the six sites that are nearest-neighbors of $i$ in the original lattice (all sites that, in the original lattice, are at a distance $r=1$ from $i$). In Fig. \ref{Fig6}, we depict the ratio between $\sigma_{\Gamma(0)}$, the variance of $\Gamma(0)$, and $\sigma_{\Gamma(1)}$, the variance of $\Gamma(1)$. This ratio does not show any spectacular behavior when the transition line is crossed.

The conclusion is that although quenched random systems are not self-averaging everywhere \cite{dsy93,wd95,ah96,wd98,mf06}, this undesirable feature can be turned into a useful tool. The strength of self-averaging expressed in terms of signal to noise ratios in quantities that in pure systems may not even be interesting, becomes a useful quantity to study. These supply additional measures of the order in the system, because, local bounded quantities lead to signal to noise ratios that depend on the correlation length in a generic way.




\end{document}